\documentclass[11pt]{paper}

\usepackage{amssymb}
\usepackage{amscd}
\usepackage{amsfonts}
\usepackage{amssymb}
\usepackage{cite}
\usepackage{amsmath}
\usepackage{epsfig}
\usepackage{graphicx, subfigure}
\usepackage{color, graphics}
\usepackage{pdfsync}
\usepackage{hyperref}
\usepackage[all,cmtip]{xy}
\usepackage{setspace}
\usepackage[all,cmtip]{xy}
\usepackage{tikz}
\usepackage{pst-all}
\usepackage{pstricks-add}
 \newtheorem{thm}{\textbf{Theorem}}[section]

 \newtheorem{rem}{\textbf{Remark}}[section]
 \numberwithin{equation}{section}

\renewcommand{\le}{\leqslant}\renewcommand{\leq}{\leqslant}
\renewcommand{\ge}{\geqslant}

\allowdisplaybreaks

\newcommand{\R}{\mathbb{ R}}

\DeclareMathOperator{\pr}{pr}
\DeclareMathOperator{\Span}{span}
\DeclareMathOperator{\diag}{diag}

\DeclareMathOperator{\tr}{tr}

\DeclareMathOperator{\ddim}{ddim}
\DeclareMathOperator{\dind}{dind}

\def\corank{\mathrm{corank}}

\def\diag{\mathrm{diag}}

\def\rank{\mathrm{rank}}

\def\px1{p_{x_1}}
\def\px2{p_{x_2}}
\def\pu1{p_{u_1}}

\begin{document}

\title{Heavy rigid body with a gyroscope in $\R^n$}

\author{Vladimir Dragovi\'c, Borislav Gaji\'c, and Bo\v zidar Jovanovi\'c}

\maketitle

\noindent{\small Department of Mathematical Sciences, The University of Texas at Dallas,  USA \& Mathematical Institute SANU, Belgrade, Serbia
\footnote{{\sc email:} Vladimir.Dragovic@utdallas.edu}}

\noindent{\small Mathematical Institute SANU, Belgrade,
Serbia\footnote{{\sc email:} gajab@mi.sanu.ac.rs}$^,$\footnote{{\sc email:} bozaj@mi.sanu.ac.rs}}

\begin{abstract}  Starting from the following multidimensional integrable generalizations of
the heavy rigid body systems: the Euler top, the Lagrange top, the
Lagrange bitop, and the totally symmetric case, we add to each of them a
gyroscope. For each of the newly constructed systems, we provide a
polynomial matrix Lax representation and prove Liouville integrability.\footnote{MSC 2020: 70E45, 70E40, 37J35, 70G65}. 
\end{abstract}

\maketitle

\section{Introduction: heavy rigid body with a gyroscope in $\R^3$}

Let us consider a motion of a rigid body in $\R^3$ around a fixed point with a gyroscope of a constant angular momentum $\mathbf L=(L_1,L_2,L_3)$ in the body frame. Let $I$ be the operator of inertia of the  system body+gyroscope (see Section \ref{sec2}) and let $\Omega$ be the angular velocity of the body in the moving frame. Then the total angular momentum of the system body+gyroscope is $\mathbf K=\mathbf M+\mathbf L$, where
$\mathbf M=I\Omega$.

Let $\gamma$ be the unit vector fixed in the space $\R^3$ in the direction of a homogeneous gravitational field and $\Gamma$ be the same vector considered in the frame attached to the body. The Euler-Poisson equations of motion in the body frame are given by
\begin{equation}\label{MEP}
\dot{\mathbf M}=\big(\mathbf M+\mathbf L\big) \times \Omega+\Gamma\times \chi, \qquad \dot\Gamma=\Gamma\times\Omega, \qquad \Omega=I^{-1}\mathbf M,
\end{equation}
where $\chi=(\chi_1,\chi_2,\chi_3)$ is the position of the center of mass of the system body+gyro\-scope, multiplied by the mass of the system $m$  and the constant $g$ of the homogeneous gravitational field.

Note that the equations \eqref{MEP} are Hamiltonian
\[
\dot M_i=\{M_i,H\}_\mathbf L, \qquad \dot\Gamma_i=\{\Gamma_i,H\}_\mathbf L,
 \]
 where the Hamiltonian is
 $$
 H=\frac12\langle I^{-1}\mathbf M,\mathbf M\rangle+\langle \chi,\Gamma\rangle,
 $$
with respect to the magnetic Poisson bracket on $\R^6(\mathbf M,\Gamma)$:
\begin{equation}\label{MPB}
\{F,G\}_\mathbf L\vert_{(\mathbf M,\Gamma)}=
-\Big\langle \mathbf M+\mathbf L,\frac{\partial F}{\partial \mathbf M}\times \frac{\partial G}{\partial \mathbf M}\Big\rangle-
\Big\langle \Gamma,\frac{\partial F}{\partial \mathbf M}\times \frac{\partial G}{\partial \Gamma}+\frac{\partial F}{\partial \Gamma}\times \frac{\partial G}{\partial \mathbf M}\Big\rangle.
\end{equation}
The magnetic Poisson bracket is obtained from the standard one by the argument translation.

It is also convenient to write these equations of motion with a Hamiltonian having a linear term.
Consider the total angular momentum $\mathbf K=\mathbf M+\mathbf L$. Then the system \eqref{MEP} is equivalent to the system
\begin{equation}\label{EP}
\dot{\mathbf K}=\mathbf K\times \Omega+\Gamma\times \chi, \qquad
\dot\Gamma=\Gamma\times\Omega, \qquad \Omega=I^{-1}\big(\mathbf K-\mathbf L\big).
\end{equation}
This system is Hamiltonian with respect to the standard Poisson bracket $\{\cdot,\cdot\}_0$ on $\R^6(\mathbf K,\Gamma)$. The standard Poisson bracket is obtained from the magnetic one by setting $\mathbf M\mapsto \mathbf K$ and $\mathbf L\mapsto 0$ in \eqref{MPB}:
\[
\dot K_i=\{K_i,H_1\}_0, \qquad \dot\Gamma_i=\{\Gamma_i,H_1\}_0, \qquad H_1=\frac12\langle\mathbf K,I^{-1}\mathbf K\rangle-\langle \mathbf K,I^{-1}\mathbf L\rangle+\langle\Gamma,\chi\rangle.
\]

As in the case of a heavy rigid body in $\mathbb R^3$ without a gyroscope,  in addition to the Hamiltonian function $H$, the geometric integral $\langle \Gamma,\Gamma\rangle=1$, and the area integral $\langle \mathbf M+\mathbf L,\Gamma\rangle$, there exists a fourth independent first integral of motion $F$ only in the three remarkable cases,  see e.g. \cite{BM2019}:

\begin{itemize}

\item\emph{The Euler top}:  $I=\diag(A,B,C)$, $\mathbf L=(L_1,L_2,L_3)$, $\chi=0$, $F=\langle \mathbf M+\mathbf L,\mathbf M+\mathbf L\rangle$, Zhukovskiy \cite{Zh} and  Volterra  \cite{Vol}.

\item\emph{The Lagrange top}:  $I=\diag(A,A,C)$, $\chi=(0,0,\chi_3)$, $\mathbf L=(0,0,\eta)$, $F=M_3$.

\item\emph{The Kowalevski top}: $I=\diag(1,1,\frac12)$, $\chi=(\chi_1,0,0)$, $\mathbf L=(0,0,\eta)$,
$
F=(M_1^2-M_2^2-2\chi_1\Gamma_1)^2+(2M_1M_2-2\chi_1\Gamma_2)^2+8\eta(M_3-2\eta)(M_1^2+M_2^2)-16\chi_1\eta M_1\Gamma_3.
$
\end{itemize}

The Kowalevski top with a gyroscope was obtained one hundred years after the celebrated Kowalevski paper \cite{Kow}, almost simultaneously in three different papers: by Yehia \cite{Y}, Komarov \cite{K}, and Reyman and  Semenov-Tian-Shansky \cite{RST}. Lax representations of the problems can be found in e.g. \cite{RaMo, D1997, RST, ST}. Recently, the dual Lax representations
in the space frame were given in \cite{Jo2025}. Reyman and  Semenov-Tian-Shansky \cite{RST, RS} also presented a multidimensional integrable generalization of
the Kowalevski top with a gyroscope.

The  main aim of the present paper is to construct multidimensional integrable versions of the Lagrange top with a gyroscope (Theorem \ref{glavna}, Section \ref{sec4} and Theorem \ref{glavna2}, Section \ref{sec5}) and the Euler top with a gyroscope (Theorem \ref{manakov-giroskop}, Section \ref{sec6}).
We also construct a totally symmetric heavy rigid body with a gyroscope in $\mathbb R^n$, that for $n=3$ and $n=4$ reduces to the Lagrange case but is essentially different from  the multidimensional Lagrange cases for $n\ge 5$ (Theorem \ref{simetricni}, Section \ref{sec4}).

\section{The heritage and motivation: works of Bobilev, Zhukovskiy, Volterra, and Demchenko}\label{sec2}

The Euler case of motion of a rigid body with a gyroscope about the fixed point is considered by Zhukovskiy in 1885, see e.g.\cite{Zh}.
He also presented an interesting geometric interpretation of the motion (see also \cite{DGJ2026}).
The equations of motion were integrated by Volterra in 1899 \cite{Vol}. For further studies, including bifurcation analysis, see, for example  \cite{Bas1, Bas2, Bas3, BM2019, Khar, GK, W}.

Let $\ell$ be the  gyroscope axis, fixed in the body, that does not need to coincide with any of the principal axes of inertia of the rigid body.
The fixed point $O$ is taken to coincide with the center of mass of the
gyroscope. We denote by $C_2$ the moment of inertia of the gyroscope with respect to its axis, and by $A_2$ the moment of inertia of the gyroscope with respect to the line that contains $O$ and lies in the plane $\sigma$, which is orthogonal  to the axis of the gyroscope. Due to the symmetry of the gyroscope,  the moment of inertia for an arbitrary line in the plane $\sigma$ that contains $O$ has the moment of inertia equal to $A_2$. It is assumed that there are no additional external forces acting on the gyroscope.
Zhukovskiy concluded that the angular velocity  $\Omega'_3$ of the gyroscope with respect to its axis is constant \cite{Zh}.
The  total angular momentum of the system body+gyroscope $\mathbf{K}$ has three parts: the angular momentum of the body $\mathbf{M}_0$ and two components of the angular momentum of the gyroscope, (i) the angular momentum $\mathbf{L}=C_2\Omega'_3\mathbf{E}'_3$ with respect to the gyroscope axis  $\ell$, and (ii) $\mathbf{L}'$ that lies in the plane $\sigma=\ell^\perp$, where $\mathbf E_3'$ is the unit vector that defines $\ell$. Zhukovskiy observed that $\mathbf{L}'$ could be included in the angular momentum of the body, by gluing to the body an infinite thin  rod lying on the gyroscope axis and having the moment of inertia equal to $A_2$ with respect to the line perpendicular to the axis of the gyroscope that belongs to the plane $\sigma$. Such a body Zhukovskiy called \emph{the transformed body}.

Thus, the transformed body has the angular momentum equal to $\mathbf{M}=\mathbf{M}_0+\mathbf{L}'=I\Omega$,  where $I$ is the new inertia operator, that we call \emph{inertia operator of the system body+gyroscope}, while the total angular momentum of the system is
\[
\mathbf K=\mathbf M+\mathbf L=I\Omega+\mathbf L.
\]
Usually, the constant angular momentum $\mathbf L$ is referred to as the \emph{angular momentum of the gyroscope}.

 Even before Volterra established the solution of the Euler case with a
gyroscope \cite{Vol}, Bobilev in 1892 solved a nonholonomic problem of a
 balanced ball, for which the geometric center $O$ coincides with the mass center, with a gyroscope rolling without slipping over a plane \cite{Bob}.
Bobilev assumed that the central ellipsoid of inertia of the ball  is rotationally symmetric
and the gyroscope axis $\ell$ coincides with the axis of symmetry of the ball.
 Let $O{\mathbf E}_1{\mathbf E}_2{\mathbf E}_3$ and $O{\mathbf E}'_1{\mathbf E}'_2{\mathbf E}'_3$ be the moving frames attached to the ball
and the gyroscope in which
the inertia operator has the forms $I_1=(A_1,A_1,C_1)$ and  $I_2=(A_2,A_2,C_2)$, respectively.
 The axis $\ell$ of the gyroscope is fixed with respect to the ball
and coincides with the axis of symmetry of the inertia ellipsoid of the ball (${\mathbf E}_3={\mathbf E}'_3$).
Then   the ball+gyroscope inertia operator $I$  and the angular momentum $\mathbf L$  for the Bobilev case are given by:
\begin{equation}\label{bobiljev}
I=\diag(A,A,C)=\diag(A_1+A_2,A_1+A_2,C_1) \quad \text{and} \quad \mathbf L=C_2\Omega'_3{\mathbf E}_3.
\end{equation}
 Bobilev  solved the equations of motion of this system in elliptic quadratures.
Zhuko\-vskiy (1893) noticed  that the reduction to the elliptic quadratures becomes simpler with an additional assumption (\emph{called the Zhukovskiy condition})
that $I$ is proportional to the identity matrix  (see \cite{Zhuk1893}), or
\begin{equation}\label{uslovZh}
C_1=A_1+A_2.
\end{equation}
In 1985 Markeev proved that the equations of motion of the general case of rolling of a balanced ball with a gyroscope can be also solved in quadratures \cite{Markeev1985}.

Demchenko, in his PhD thesis in 1923, used the Zhukovskiy condition \eqref{uslovZh} to integrate the equations of motion of the problem of rolling of a gyroscopic ball over a sphere \cite{Dem1924} (see also \cite{DGJ}).
The integrability  of equations of motion of the problem of rolling of the gyroscopic ball over a sphere with the Bobilev conditions \eqref{bobiljev} was proved by Borisov, Mamaev and Kilin, see \cite{BoMa}.
The question about the existence of an integrable case of a dynamically nonsymmetric ball with a gyroscope rolling over a sphere is still open.

In \cite{DGJ2023}, we considered an $n$--dimensional version of the Demchenko case: a nonholonomic problem of rolling of a ball with a gyroscope without slipping and twisting over a hyperplane and over a sphere in $\R^n$, where the inertia operator of the system ball+gyroscope is proportional to the identity operator.
It is an example of a $G$-Chaplygin system with gyroscopic forces and allows for a reduction to the magnetic geodesic flow on the sphere $S^{n-1}$ placed in the homogeneous magnetic field in $\R^n$. It appeared that this magnetic flow, although quite simple when considered in $\R^n$, after being restricted to a sphere, provides a new integrable model closely related to the Neumann system \cite{BKM2025, DGJ2025, DGJ2025b}.

This motivates our present study of  higher-dimensional rigid bodies with a gyroscope.

\section{Two $n$-dimensional generalizations}\label{sec3}

The standard Poisson bracket on $\R^6(\mathbf M,\Gamma)$ can be seen both as the standard Lie-Poisson bracket
on the dual space of the Lie algebra of Euclidean motions $e(3)$ and on the dual space of the semi-direct product $so(3)\times so(3)$. Thus, there are
two natural higher-dimensional generalizations of
the Euler-Poisson equations of motion of a heavy rigid-body in ${\mathbb R^3}$. The first generalization is on $(so(n)\times so(n))^*$ (see Ratiu \cite{Ra2}) and  the second one is on
$e(n)^*=(so(n)\times \R^n)^*$ (see Belyaev \cite{Be}).

Reyman and  Semenov-Tian-Shansky  used the second generalization in the presence of $q$ homogeneous fields, with $q\le n$,  to obtain multidimensional integrable generalizations of
the Kowalevski top. The equations of motion of those systems are
the Euler-Poisson equations on  $\big(so(n)\times (\R^n)^q\big)^*$. They modelled a gyroscope by adding a linear in momenta term in the Hamiltonian function for $q=2$ (see \cite{RST, RS}).

Note that a heavy rigid body with a gyroscope on the full phase space $T^*SO(n)$ is described as a Hamiltonian system with the standard symplectic structure perturbed by the exact two-form on $SO(n)$ (see e.g. \cite{novikov}).  The magnetic systems on Lie groups with respect to non-exact magnetic $2$-forms were studied in \cite{MS, T}.

\subsection{Rigid body systems on $(so(n)\times so(n))^*$}

By the analogy with \eqref{MEP}, the equations of motion of a heavy $n$-dimensional rigid body on $(so(n)\times
so(n))^*(M,\Gamma)$ with a gyroscope are:
\begin{equation}
\dot M=[M+L, \Omega]+[\Gamma, \chi],\qquad \dot\Gamma=[\Gamma,\Omega], \qquad \Omega=I^{-1}M,
\label{rMEP}
\end{equation}
where $M, \Omega, \Gamma,\chi, L\in so(n)$, and $\chi$ and $L$
are some constant matrices. Here $I\colon so(n)\to so(n)$ is
 the  body+gyroscope inertia operator:
\begin{equation}\label{MANAKOV}
M=J\Omega+\Omega J,
\end{equation}
where $J=\diag(J_1,\dots,J_n)$ is the mass tensor  of the system body+gyroscope
in the orthonormal basis $[\mathbf E_1,\dots,\mathbf E_n]$ of $\R^n$ formed by principal axes of inertia of the rigid body (see \cite{FK1995}  and Section \ref{sec2}).
The operator \eqref{MANAKOV} belongs to the class of the Manakov operators \cite{Ma} on $so(n)$.

The Euler-Poisson equations \eqref{rMEP} are Hamiltonian
with respect to the magnetic Lie-Poisson structure on
$(so(n)\times so(n))^*\cong so(n)\times so(n)$:
\[
\{M_{ij}, M_{jk}\}_L=-M_{ik}-L_{ik},\qquad \{M_{ij},
\Gamma_{jk}\}_L=-\Gamma_{ik},\qquad
\{\Gamma_{ij},\Gamma_{kl}\}_L=0,
\]
with the
Hamiltonian function
\begin{equation*}
\label{hf-ha}
H=\frac12\langle
M,\Omega\rangle+\langle\chi,\Gamma\rangle=-\frac14\tr(M\Omega)-\frac12\tr(\chi\Gamma),
\end{equation*}

Here, we identified the dual spaces by the invariant scalar product on the \emph{direct} product $so(n)\times so(n)$:
\[
\langle(\xi_1,\eta_1),(\xi_2,\eta_2)\rangle=\langle \xi_1,\xi_2\rangle+\langle
\eta_1,\eta_2\rangle=-\frac12\tr(\xi_1\xi_2)-\frac12\tr(\eta_1\eta_2), \,\, (\xi_i,\eta_i)\in so(n)\times so(n).
\]

Equivalently, by setting $K=M+L$,  the system \eqref{rMEP} takes the form:
\begin{equation}\label{rEP}
\dot K=[K, \Omega]+[\Gamma, \chi],\qquad
\dot\Gamma=[\Gamma,\Omega], \qquad \Omega=I^{-1}\big(K-L\big).
\end{equation}

The system \eqref{rEP} is Hamiltonian with respect to the standard Poisson brackets on $(so(n)\times so(n))^*$:
\[
\{K_{ij}, K_{jk}\}_0=-K_{ik},\qquad \{K_{ij},
\Gamma_{jk}\}_0=-\Gamma_{ik},\qquad \{\Gamma_{ij},\Gamma_{kl}\}_0=0,
\]
with the Hamiltonian function  having a linear in momenta term
\begin{equation}\label{eq:H1}
H_1=\frac12\langle K,I^{-1} K\rangle-\langle  K,I^{-1} L\rangle+\langle\Gamma,\chi\rangle.
\end{equation}

The mapping $M\mapsto K=M+L$ is a Poisson mapping between the magnetic and the standard Lie-Poisson brackets. Therefore,
the Casimir functions of the standard structure
\[
P_k=\tr(\Gamma^{2k}), \qquad Q_k=\tr(K\Gamma^{2k-1}), \qquad k=1,\dots,\Big[\frac{n}2\Big],
\]
by setting $K=M+L$, induce the Casimir functions of the magnetic structure.
In both cases, the dimension of a generic symplectic leaf is
${n}(n-1)-2\Big[\frac n2\Big]$.

\subsection{Rigid body systems on $e(n)^*$}

In the second model, the magnetic Euler--Poisson equations are:
\begin{equation}\label{bMEP}
\dot M =[M+L,\Omega ]+\chi \wedge \Gamma, \qquad \dot \Gamma = -\Omega\cdot \Gamma, \qquad \Omega=I^{-1}M,
\end{equation}
where $\chi\in\R^n$ is
the vector of the mass centre, $\Gamma\in\R^n$ is the vertical vector
considered in the moving coordinate system, and
$L$ is the angular momentum of the gyroscope.
The equations of motion are Hamiltonian with respect to the magnetic Poisson structure on $e(n)^*(M,\Gamma)$:
\[
\{M_{ij},M_{jk}\}_L=-M_{ik}-L_{ik}, \qquad \{\Gamma_i,\Gamma_j\}_L=0,
\qquad \{M_{ij},\Gamma_k\}_L=-\Gamma_i
\delta_{jk}+\Gamma_j\delta_{ik},
\]
with the Hamiltonian function:
\[
H=\frac12\langle
M,\Omega\rangle+\langle\chi,\Gamma\rangle=-\frac14\tr(M\Omega)+\sum_{i=1}^n \chi_i \Gamma_i.
\]

Here we identified  $e(n)^*\cong e(n)$ by the use of the
non-invariant scalar product:
\[
\langle(\xi_1,\eta_1),(\xi_2,\eta_2)\rangle=\langle \xi_1,\xi_2\rangle+\langle
\eta_1,\eta_2\rangle=-\frac12\tr(\xi_1\xi_2)+\sum_{i=1}^n \eta_{1,i}\eta_{2,i}, \, (\xi_i,\eta_i)\in so(n)\times \R^n.
\]

By setting $K=M+L$, the system of equations of motion \eqref{bMEP} gets the form:

\begin{equation}\label{bEP}
\dot K=[K, \Omega]+\chi\wedge\Gamma ,\qquad
\dot\Gamma=-\Omega\cdot \Gamma, \qquad \Omega=I^{-1}\big(K- L\big).
\end{equation}

The system \eqref{bEP} is Hamiltonian with respect to the standard Poisson brackets on $e(n)^*$:
\[
\{K_{ij}, K_{jk}\}_0=-K_{ik},\qquad \{\Gamma_i,\Gamma_j\}_0=0,
\qquad \{K_{ij},\Gamma_k\}_0=-\Gamma_i
\delta_{jk}+\Gamma_j\delta_{ik},
\]
with the Hamiltonian function with a linear term:
\[
H_1=\frac12\langle K,I^{-1} K\rangle-\langle  K,I^{-1} L\rangle+\langle\Gamma,\chi\rangle.
\]

In the description of the Casimir functions, we follow Brailov's construction described in \cite{TF}.
The standard Lie-Poisson bracket on $e(n)^*(K,\Gamma)$ can be seen as a contraction of the standard Lie-Poisson bracket $\{\cdot,\cdot\}_1$ on
$so(n+1)^*(\hat K)$ as follows. Set
\[
\hat K_{ij}=K_{ij}, \qquad \hat K_{i,n+1}=\Gamma_i, \qquad i,j=1,\dots n,
\]
and define for each $\lambda$ a Poisson structure $\{\cdot,\cdot\}_\lambda$ and also the limit Poisson structure $\{\cdot,\cdot\}_0$ obtained in the limit when  $\lambda\to 0$ on $so(n+1)^*$ by
\begin{equation}\label{contraction}
\begin{aligned}
&\{\hat K_{ij},\hat K_{jk}\}_{\lambda}=-\hat K_{ik}=-K_{ik}=\{K_{ij},K_{jk}\}_0, \qquad i,j,k=1,\dots,n,\\
&\{\Gamma_i,\Gamma_j\}_{\lambda}=-\{\hat K_{i,n+1},\hat K_{n+1,j}\}_{\lambda}=\lambda\hat K_{ij}, \,\{\Gamma_i,\Gamma_j\}_0=0,\\
& \{\hat K_{ij},\hat K_{k,n+1}\}_{\lambda}=-\delta_{jk}\hat K_{i,n+1}+\delta_{ik}\hat K_{j,n+1}=-\delta_{jk}\Gamma_i+\delta_{ik}\Gamma_j=\{K_{ij},\Gamma_{k}\}_0.
\end{aligned}
\end{equation}

Therefore, the Casimir functions on $(e(n)^*,\{\cdot,\cdot\}_0)$ can be also seen as contractions of the invariants $p_{k}$ on $(so(n+1)^*,\{\cdot,\cdot\}_1)$, which we take in the form:
\begin{align*}
p_{k}(\hat K)=\sum_{1\le i_1<\dots<i_{2k}\le n+1}\det(\hat K^{i_1,\dots,i_{2k}}_{i_1,\dots,i_{2k}}), \qquad k=1,\dots,\Big[\frac{n+1}2\Big],
\end{align*}
 where $\hat K^{i_1,\dots,i_{2k}}_{i_1,\dots,i_{2k}}$ is a ($2k\times 2k$) skew-symmetric matrix $\hat K_{ij}$, $i,j\in \{i_1,\dots,i_{2k}\}$.

By definition (see \cite{TF}), the contraction of $p_k(\hat K)$ is the polynomial $q_k(K,\Gamma)$ consisting of the terms with the maximal degree in $\Gamma_i=\hat K_{i,n+1}$.
As a result, we get the invariants
\begin{align*}
q_{k}(K,\Gamma)=\sum_{1\le i_1<\dots<i_{2k-1}<n+1}\det(\hat K^{i_1,\dots,i_{2k-1},n+1}_{i_1,\dots,i_{2k-1},n+1}), \qquad k=1,\dots,\Big[\frac{n+1}2\Big],
\end{align*}
and the dimension of a generic symplectic leaf is
$\frac{(n+1)(n-1)}2-\Big[\frac {n+1}2\Big]$.
For example
\begin{align*}
q_1=&\sum_{i=1}^n \det(\hat K^{i,n+1}_{i,n+1})   =\sum_{i=1}^n \Gamma_i^2=\langle \Gamma,\Gamma\rangle, \\
q_2=&\sum_{1\le i<j<k<n+1}\det(\hat K^{i,j,k,n+1}_{i,j,k,n+1})=\sum_{1\le i<j<k<n+1}\big( -K_{ij}\Gamma_k+K_{ik}\Gamma_j-K_{jk}\Gamma_i\big)^2.
\end{align*}
The invariants of the magnetic brackets are $q_k(K,\Gamma)\vert_{K=M+L}$.
Another descriptions of the invariants can be found in \cite{Vor, Zhd}.

\section{The Lagrange and the totally symmetric heavy rigid body with a gyroscope on $(so(n)\times so(n))^*$}\label{sec4}

\subsection{The Lagrange top and the Lagrange bitop}

Let $[\mathbf E_1,\dots,\mathbf E_n]$ be the standard orthonormal basis of $\R^n$.
In \cite{Ra2} the $n$-dimensional \emph{Lagrange top}  on $(so(n)\times so(n))^*$ was constructed.
Starting from the mass tensor $J$ and a matrix $\chi$ given by
\[
J=\diag(J_1,\dots,J_n), \quad J_1=J_2=\alpha_1, \quad J_3=\dots=J_n=\alpha_2, \quad \chi=\chi_{12}\mathbf E_1\wedge\mathbf E_2,
\]
the Hamiltonian of the $n$-dimensional Lagrange top  on $(so(n)\times so(n))^*$ was defined by
\begin{equation}\label{Hl}
H_{l}=\frac12\Big(\frac1{2\alpha_1}M_{12}^2+\frac1{\alpha_1+\alpha_2}\sum_{p=3}^{n}(M_{1p}^2+M_{2p}^2)+\frac1{2\alpha_2}\sum_{3\leq p<q\leq n}M_{pq}^2\Big)+\chi_{12}\Gamma_{12}.
\end{equation}
For $n=3$, the above Hamiltonian coincides  with the usual Hamiltonian of the Lagrange top with a gyroscope heaving the inertia operator
$I=\diag(\alpha_1+\alpha_2,\alpha_1+\alpha_2,2\alpha_1)$ and
the position of the center of mass of the system body+gyroscope, multiplied by the mass of the system $m$  and the gravitational constant $g$ given by
$\chi=(0,0,\chi_{12})$.

Consider the subalgebra
\[
\mathfrak h=\Span\{\mathbf E_1\wedge \mathbf E_2, \mathbf E_p\wedge \mathbf E_q\,\vert\, 3\leq p<q\leq n\}\cong so(2)\oplus so(n-2).
\]
For $\alpha_1\ne \alpha_2$, it coincides with the isotropy subalgebra $so(n)_J=\{\xi\in so(n)\,\vert\, [\xi,J]=0\}$.
Let $\mathfrak v$ be the orthogonal complements of $\mathfrak h$ within $so(n)$ with respect to the invariant scalar product $\langle\cdot,\cdot\rangle$:
\[
\mathfrak v=\Span \{\mathbf E_p\wedge \mathbf E_q\,\vert\, 1\le p\le2, 3\le q\le n\}.
\]

Then $I$ preserves the decomposition $so(n)=\mathfrak h\oplus \mathfrak v=so(2)\oplus so(n-2)\oplus\mathfrak v$,
\[
 I=2\alpha_1\pr_{so(2)}+2\alpha_2\pr_{so(n-2)}+(\alpha_1+\alpha_2)\pr_\mathfrak v,
\]
and $\chi$ is a central element in $\mathfrak h$:
\begin{equation}\label{central}
[\chi,\mathfrak h]=0.
\end{equation}

In dimension $n=4$, there is also an additional integrable case (constructed in \cite{DrGa2001}, where a Lax representation was also provided, see also \cite{DrGa2004} for a detailed algebro-geometric integration),  known as \emph{the Lagrange bitop}, defined by
\[
J=\diag(\alpha_1,\alpha_1,\alpha_2,\alpha_2), \qquad \chi=\chi_{12}\mathbf E_1\wedge\mathbf E_2+\chi_{34} \mathbf E_3\wedge \mathbf E_4.
\]
In the case of the Lagrange bitop, the algebra $\mathfrak h$ is commutative, which allows for $\chi$ to be an arbitrary element in $\mathfrak h$ with the property \eqref{central}.
The Hamiltonian of the Lagrange bitop is
\begin{equation}\label{Hb}
H_{lb}=\frac12\Big(\frac1{2\alpha_1}M_{12}^2+\frac1{\alpha_1+\alpha_2}(M_{13}^2+M_{23}^2+ M_{14}^2+M_{24}^2)+\frac1{2\alpha_2}M_{34}^2\Big)+\chi_{12}\Gamma_{12}+\chi_{34}\Gamma_{34}.
\end{equation}

We now define both the Lagrange top and the Lagrange bitop with a gyroscope, assuming in addition that $L$ is an arbitrary element of $\mathfrak h$:
\begin{equation}\label{L.L}
L=L_{12}\mathbf E_1\wedge \mathbf E_2+\sum_{3\leq p<q\leq n} L_{pq} \mathbf E_p\wedge \mathbf E_q.
\end{equation}

The following theorem discusses integrability of these new systems.

\begin{thm}\label{glavna}
(i) The equations of motion of a heavy rigid body with the gyroscope \eqref{rMEP} in the cases of
the Lagrange top \eqref{Hl} and the Lagrange bitop \eqref{Hb} with the gyroscope angular momentum \eqref{L.L},
are equivalent to the
polynomial matrix Lax representations:
\begin{equation}
\dot {\mathbb L}(\lambda)=[{\mathbb L}(\lambda), {\mathbb A}(\lambda)],\quad {\mathbb L}(\lambda)=\Gamma+ \lambda(M+L)+{\lambda^2}({\alpha_1+\alpha_2})\chi,\quad {\mathbb A}(\lambda)=\Omega+\lambda\chi.
\label{matrix}
\end{equation}

(ii) For a generic $L\in\mathfrak h$, the Lagrange top and the Lagrange bitop with a gyroscope are Liouville integrable.
\end{thm}

\noindent\emph{Proof.}
(i) The polynomial matrix equations \eqref{matrix} in the first and the zero degree in $\lambda$ are equivalent to the first and the second equation in \eqref{rMEP}, respectively:
\begin{align*}
&\lambda^1\colon \qquad \dot M=[M+L,I^{-1}M]+[\Gamma,\chi],\\
&\lambda^0\colon\qquad   \dot\Gamma=[\Gamma,I^{-1}M].
\end{align*}
For the second degree  in $\lambda$, we get:
\[
\lambda^2\colon\qquad \frac{d}{dt}\big(({\alpha_1+\alpha_2})\chi\big)=[({\alpha_1+\alpha_2})\chi,\Omega]+[M+L,\chi]=[({\alpha_1+\alpha_2})\chi,\Omega]+[M,\chi]=0,
\]
where we  use
$[\chi,L]=0$ (see \eqref{central}).
The last identity was proved in \cite{Ra2} for the Lagrange top and in \cite{DrGa2004} for the Lagrange bitop.

(ii) To prove the complete integrability, we set $K=M+L$, consider the standard Lie-Poisson bracket $\{\cdot,\cdot\}_0$
and  the gyroscopic system \eqref{rEP} with the Hamiltonian $H_1$ (see \eqref{eq:H1})
obtained from \eqref{Hl} and \eqref{Hb}:
\[
H_1(K,\Gamma)=H_l(M,\Gamma)\vert_{M=K-L}, \qquad  (H_1(K,\Gamma)=H_{lb}(M,\Gamma)\vert_{M=K-L}).
\]
Together with \eqref{rEP}, let us consider the standard Lagrange top as well as the Lagrange bitop:
\begin{equation*}\label{LEP}
\dot K=[K, \Omega]+[\Gamma, \chi], \qquad \dot\Gamma=[\Gamma,\Omega], \qquad \Omega=I^{-1}K,
\end{equation*}
with the Hamiltonian $H_0(K,\Gamma)=H_l(M,\Gamma)\vert_{M=K}$ ($H_0(K,\Gamma)=H_{lb}(M,\Gamma)\vert_{M=K}$).

Due to $SO(2)\times SO(n-2)$-symmetry,
the projection of the angular momentum $K$ to $\mathfrak h$ is conserved:
\[
\frac{d}{dt}K_\mathfrak h=0,
\]
that is, the Lagrange tops, as well as the Lagrange bitop for $n=4$, have Noether first integrals forming the set of linear functions on $\mathfrak h$:
\[
\mathcal S=\Span\{K_{12}, K_{pq}\, \vert\,  3\le p<q<n\}.
\]

Let $f_1,\dots,f_\ell$ be independent polynomial first integrals obtained from the Lax representation, when we set $M=K$ and $L=0$ in \eqref{matrix}.
These first integrals are the coefficients  with a degree of $\lambda$ of the polynomials
\[
\tr\big((\Gamma+\lambda K+{\lambda^2}({J_1+J_3})\chi)^{2k}\big), \qquad k=1,\dots,\Big[\frac{n}2\Big].
\]

The first integrals $f_i$ Lie-Poisson commute between themselves and with the
Noether first  integrals:
\begin{equation}\label{r1}
\begin{aligned}
&\{H_0,f_i\}_0=0, \quad \{H_0,K_{12}\}_0=0, \quad \{H_0,K_{pq}\}_0=0,  \qquad i,j=1,\dots,\ell,   \\
&\{f_i,f_j\}_0=0, \quad\, \{K_{12}, f_i\}_0=0, \qquad \{K_{pq},f_i\}_0=0,  \qquad 3\le p<q<n,
\end{aligned}
\end{equation}
providing non-commutative integrability (see \cite{MF2, N}) of the Lagrange top for $n\ge 5$ and commutative
integrability (or Liouville integrability, see Arnold \cite{Ar})  of the Lagrange bitop (see Theorem \ref{noncommutative} below) on generic symplectic leaves within $(so(n)\times so(n))^*$.

We have
\begin{equation}\label{r2}
H_1(K,\Gamma)=H_0(K,\Gamma)-H_\mathfrak h(K),
\end{equation}
where
\[
H_\mathfrak h(K)=\frac{1}{2\alpha_1}L_{12} K_{12}+\frac{1}{2\alpha_2}\sum_{3\leq p<q\leq n} L_{pq} K_{pq}=\langle I^{-1}L,K_\mathfrak h\rangle.
\]
By $K_\mathfrak h$, we denote the projection of $K$ to $\mathfrak h$ with respect to the decomposition $so(n)=\mathfrak h\oplus\mathfrak v$.

Let us consider the linear Hamiltonian system with the Hamiltonian $H_\mathfrak h$ on the Lie-Poisson space $(\mathfrak h,\{\cdot,\cdot\}_0)$:
\[
\dot K_\mathfrak h=[K_\mathfrak h,I^{-1}L].
\]

It has the following simple matrix representation
\[
\dot {\mathbb L}_\mathfrak h(\lambda)=[{\mathbb L}_\mathfrak h(\lambda), {\mathbb A}_\mathfrak h],\qquad
{\mathbb L}_\mathfrak h(\lambda)=K_\mathfrak h+\lambda I^{-1} L,\qquad {\mathbb A}_\mathfrak h=I^{-1}L.
\]

Thus, the coefficients $q_k(K_\mathfrak h)$ in $\lambda^k$ of the polynomials
\[
\tr\big((K_\mathfrak h+\lambda I^{-1}L)^{2i}\big)=\sum_k q_k(K_\mathfrak h)\lambda^k, \qquad i=1,\dots,\Big[\frac{n}2\Big],
\]
 are first integrals of the Hamiltonian system:
\begin{equation}\label{r3}
\{H_\mathfrak h,q_k\}_0=0.
\end{equation}

Moreover, according to the Mishchenko-Fomenko argument translation method, the polynomials are in involution:
\begin{equation}\label{r4}
\{q_k,q_l\}_0=0,
\end{equation}
and for a generic $L\in \mathfrak h$, form a complete commutative set on $(\mathfrak v,\{\cdot,\cdot\}_0)$ (see \cite{MF1}).
Thus, since polynomials $q_k$ are functions of Noether first integrals $\mathcal S$,
according to \eqref{r1}, \eqref{r2}, \eqref{r3}, and \eqref{r4}, the polynomials $f_i,  q_k$
provide complete commutative integrability of the gyroscopic system:
\[
\{H_1, f_i\}_0=0, \quad \{H_1, q_k\}_0=0, \quad \{f_i,f_j\}_0=0, \quad \{q_k,q_l\}_0=0, \quad \{f_i,q_k\}_0=0.
\]
\hfill$\Box$

\begin{rem}
Apart from the Euler, Lagrange and Kowalevski integrable cases of a motion of a heavy rigid body about a fixed point,
there are also so-called partially integrable cases of such a motion, where, instead of an additional first integral, these exists an invariant relation. A famous such an example is the Hess-Appel'rot case. The Hess-Appel'rot case with gyroscope was first considered by Sretensky in 1963 in \cite{Sr1963}. In \cite{DrGa2001}, the Lax representation for this system was constructed. Starting from \cite{DrGa2001, DrGa2006, DGJ2009b}, a natural  study of the $n$-dimensional Hess-Appel'rot systems with a gyroscope can be pursued.
\end{rem}

\subsection{The totally symmetric case}

If $\alpha_1=\alpha_2$, then $M$ and $\Omega$ are proportional and $[M,\Omega]=0$.
Further,  the relation $[2\alpha_1\chi,\Omega]+[M,\chi]=0$ holds for an arbitrary $\chi\in so(n)$.
Now, let
\[
\mathfrak h=so(n)_\chi=\{\xi\in so(n)\,\vert\, [\xi,\chi]=0\}.
\]

If $\chi$ is a regular element, then $\mathfrak h$ is commutative and:
\[
\mathfrak h=\Span\{\chi,\chi^3,\dots,\chi^{2[n/2]-1}\}, \qquad \dim\mathfrak h=\Big[\frac{n}2\Big]=\rank(so(n)).
\]

\begin{thm}\label{simetricni}
(i) Let $\chi\in so(n)$ and $L\in \mathfrak h=so(n)_\chi$. The equations of motion of a heavy rigid body with the gyroscope \eqref{rMEP} in
the totally symmetric case, given with the Hamiltonian
\[
H_{sym}=\frac1{4\alpha_1}\langle M,M\rangle+\langle \Gamma,\chi\rangle,
\]
with a gyroscope having the angular momentum $L$
are equivalent to the
matrix polynomial Lax representation
\begin{equation}
\dot {\mathbb L}(\lambda)=[{\mathbb L}(\lambda), {\mathbb A}(\lambda)],\quad {\mathbb L}(\lambda)=\Gamma+ \lambda(M+L)+2{\lambda^2}\alpha_1\chi,\quad {\mathbb A}(\lambda)=\Omega+\lambda\chi.
\label{matrix2}
\end{equation}

(ii) For a generic $L\in\mathfrak h$, the heavy symmetric top with a gyroscope is Liouville integrable.
\end{thm}

Note that $\Omega=2\alpha_1 M$ holds in any orthonormal basis of $\R^n$. Thus, for $n=3$ and $n=4$, by taking suitable orthonormal basis of $\R^3$ and $\R^4$, we can always assume that $\chi$ is of the form $\chi=\chi_{12}\mathbf E_1\wedge \mathbf E_2$ and $\chi=\chi_{12}\mathbf E_1\wedge \mathbf E_2+\chi_{34}\mathbf E_3\wedge \mathbf E_4$.
Thus, for $n=3$ and $n=4$, the totally symmetric case is a special case of the Lagrange top and the Lagrange bitop. However, for $n\ge 5$ we obtain an
essentially new integrable gyroscopic model.

\

\noindent\emph{Proof.}
The proof of item (i) is the same as in Theorem \ref{glavna}.
For item (ii) we also set $K=M+L$, and consider the gyroscopic system \eqref{rEP} with respect to the standard Lie-Poisson bracket $\{\cdot,\cdot\}_0$
and with the Hamiltonian  (see \eqref{eq:H1})
\[
 H_{1}=\frac1{4\alpha_1}\langle K,K\rangle-\frac1{2\alpha_1}\langle K,L\rangle+\langle \Gamma,\chi\rangle.
\]
The rest of the proof is the same as the proof of item (ii) in Theorem \ref{glavna}, based on the Theorem \ref{noncommutative} given below.
\hfill $\Box$

\

Let us consider a heavy totally symmetric rigid body
\begin{equation}\label{symHRB}
\dot K=[\Gamma, \chi], \qquad
\dot\Gamma=[\Gamma,\Omega], \qquad \Omega=\frac{1}{2\alpha_1}K,
\end{equation}
with the Hamiltonian
\[
H_{0}=\frac1{4\alpha_1}\langle K,K\rangle+\langle \Gamma,\chi\rangle.
\]

Since $\pr_\mathfrak h[\Gamma,\chi]=0$, we have the conservation of the projection of the angular momentum $K$ to $\mathfrak h$:
\[
\frac{d}{dt}K_\mathfrak h=0.
\]

Let $\mathcal S$ be the set of linear functions on $\mathfrak h$ and let $f_1,\dots,f_\ell$ ($\ell=\ell(\chi)$) be independent first integrals from the Lax representation (set $M=K$ and $L=0$ in \eqref{matrix2}). The first integrals $f_i$ Lie-Poisson commute between themselves (see \cite{Ra2}) and with the
Noether first integrals:
\begin{equation*}\label{s1}
\{f_i,f_j\}_0=0, \qquad \{s, f_i\}_0=0, \qquad i,j=1,\dots,\ell(\chi), \qquad s\in\mathcal S,
\end{equation*}
implying complete non-commutative integrability \cite{MF2, N} of a heavy symmetric rigid body \eqref{symHRB}
on generic symplectic leaves within $(so(n)\times so(n))^*$ (see Theorem \ref{noncommutative}).

The following theorem is a special case of  Bolsinov's Theorem 1.5 in \cite{Bo}.

\begin{thm}\label{noncommutative}
Consider the set of linear functions $\mathcal S$ on $\mathfrak h=so(n)_\chi$ and the polynomials
\[
\mathcal P=\Big\{\tr\big((\Gamma+\lambda K+{\lambda^2}\alpha\chi)^{2k}\big)\,\vert\, \lambda\in\R, \, k=1,\dots,\Big[\frac{n}2\Big]\Big\},
\]
where $\alpha\ne 0$ is a real parameter. Then

(i) $\{\mathcal S,\mathcal P\}_0=0$, $\{\mathcal P,\mathcal P\}_0=0$.

(ii) $\mathcal S + \mathcal P$ is a complete set of functions on $(so(n)\times so(n))^*$:
\[
\ddim(\mathcal S+\mathcal P)+\dind(\mathcal S+\mathcal P)=\dim(so(n)\times so(n))+\corank(\{\cdot,\cdot\}_0).
\]
 \end{thm}

When $\chi$ is regular then $\mathcal S$ is commutative set, functionally dependent on $f_1,\dots,f_\ell$. In this case, Theorem \ref{noncommutative}
is proven by Ratiu \cite{Ra2}. The singular case $\chi=\chi_{12}\mathbf E_1\wedge \mathbf E_2$ was also considered in \cite{Ra2}.

\section{The Lagrange top with a gyroscope on $e(n)^*$}\label{sec5}

In \cite{Be}, Belyaev considered the $n$-dimensional \emph{Lagrange
top} on $e(n)^*$ defined by the mass tensor $J$ and the vector $\chi\in\R^n$ given by
\[
J=\diag(J_1,\dots,J_n), \quad J_1=\dots=J_{n-1}=\alpha_1, \quad J_n=\alpha_2, \quad  \chi=\chi_n\mathbf E_n.
\]

 We take $J$ as the body+gyroscope mass tensor  and consider the subalgebra
\[
\mathfrak h=\Span\{\mathbf E_p\wedge \mathbf E_q\,\vert\, 1\leq p<q\leq n-1\}\cong so(n-1).
\]
For $\alpha_1\ne \alpha_2$, it coincides with the isotropy subalgebra $so(n)_J=\{\xi\in so(n)\,\vert\, [\xi,J]=0\}$.
Let $\mathfrak v$ be the orthogonal complements of $\mathfrak h$ within $so(n)$ with respect to the invariant scalar product $\langle\cdot,\cdot\rangle$:
\[
\mathfrak v=\Span \{\mathbf E_p\wedge \mathbf E_n\,\vert\, 1\le p \leq n-1\}.
\]

Then, like in the Lagrange top on $(so(n)\times so(n))^*$,  the inertia operator $I$ preserves the decomposition $so(n)=\mathfrak h\oplus \mathfrak v$,
\[
I=2\alpha_1\pr_\mathfrak h+(\alpha_1+\alpha_2)\pr_\mathfrak v,
\]
and the Hamiltonian function takes the form
\begin{equation}\label{HLB}
H_{l}=\frac12\Big(\frac1{2\alpha_1}\sum_{1\leq p<q\leq n-1}M_{pq}^2 M_{12}^2+\frac1{\alpha_1+\alpha_2}\sum_{p=1}^{n-1}M_{pn}^2\Big)+\chi_{n}\Gamma_{n}.
\end{equation}

 For $n=3$, the Hamiltonian becomes the usual Hamiltonian of the Lagrange top with the inertia operator
$I=\diag(\alpha_1+\alpha_2,\alpha_1+\alpha_2,2\alpha_1)$ and
the position of the center of mass of the system body+gyroscope, multiplied by the mass of the system $m$  and the gravitational constant $g$ given by
$\chi=(0,0,\chi_{12})$.

As above, we define the Lagrange top with a gyroscope additionally assuming that $L$ is an arbitrary element of $\mathfrak h$:
\begin{equation}\label{L.B}
L=\sum_{1\leq p<q\leq n-1} L_{pq} \mathbf E_p\wedge \mathbf E_q.
\end{equation}

Related to the contraction \eqref{contraction}, for a given $(\xi,\eta)\in e(n)^*\cong e(n)$, we define $\hat \xi, \hat\eta\in
so(n+1)$ by:
\[
\hat \xi= \left(
\begin{array}{cc}
\xi & 0  \\
0 & 0
\end{array} \right), \quad
\hat\eta= \left( \begin{array}{cc}
{0} & \eta  \\
-\eta^t & 0
\end{array} \right).
\]

\begin{thm}\label{glavna2}
(i) The equations of motion of a heavy rigid body with the gyroscope \eqref{bMEP} in the case of
the Lagrange top \eqref{HLB} with a gyroscope with the angular momentum \eqref{L.B} is equivalent to the
polynomial matrix Lax representation
\begin{equation*} \dot {\mathbb L}(\lambda)=[{\mathbb L}(\lambda), {\mathbb A}(\lambda)],\quad
{\mathbb L}(\lambda)=\hat\Gamma+\lambda  (\hat M+\hat L)+ \lambda^2
(\alpha_1+\alpha_2)\hat\chi, \quad {\mathbb A}(\lambda)=\hat\Omega+ \lambda
\hat\chi. \label{LA_pair}
\end{equation*}

(ii) For a generic $L\in\mathfrak h$, the Lagrange top  with a gyroscope is Liuville integrable.
\end{thm}

Belyaev proved noncommutative integrability of the system without the gyroscopic term
\cite{Be}. For $L=0$, the Lax representation, as the Lax representation \eqref{matrix2}, belongs to the class of Lax matrices related to symmetric pairs given by Reyman and Semenov-Tian-Shansky \cite{RS}. The proof of Theorem \ref{glavna2} follows literarily the same lines of the proof of Theorem \ref{glavna}, with
the Noether first integrals on $\mathfrak h=so(2)\oplus so(n-2)$ being replaced by the Noether first integrals on $\mathfrak h=so(n-1)$, and using the fact that
\[
[\hat L,\hat \chi]=0.
\]

Note that here the totally symmetric case $\alpha_1=\alpha_2$ is the same as the Lagrange case, since for a generic $\chi\in\R^n$, we can choose an othonormal basis
$[\mathbf E_1,\dots,\mathbf E_n]$
of $\R^n$, such that $\chi=\chi_n\mathbf E_n$.

\section{The Euler-Manakov top with a gyroscope}\label{sec6}

Consider a motion of a free rigid body  about a fixed point with a gyroscope with the angular momentum $L$ in the body frame:
\begin{equation*}\label{ManG}
\dot M=[M+L,\Omega], \qquad \Omega=I^{-1}M,
\end{equation*}
where the inertia operator  $I$ is given by \eqref{MANAKOV}.
The equations of motion are Hamiltonian with respect to the magnetic Lie-Poisson bracket on $so(n)^*$:
\[
\{M_{ij}, M_{jk}\}_L=-M_{ik}-L_{ik}.
\]

 Following \cite{DGJ2009}, we assume a
symmetric system rigid body+gyroscope with the mass tensor $J=\diag(J_1,\dots,J_n)$ satisfying:
\begin{equation}\label{symJ}
J_1=\dots=J_{l_1}=\alpha_1,\, \dots, \, J_{n+1-l_p}=\dots=J_n=\alpha_{p},
\end{equation}
where $l_1+l_2+\dots+l_{p}=n$, $\alpha_i\ne \alpha_j$, $i\ne j$.

Let $\mathfrak h$ be the isotropy subalgebra
\begin{align*}
\mathfrak h=so(n)_J=\{\xi\in so(n)\,\vert\, [\xi,J]=0\}\cong so(l_1)\oplus so(l_2)\oplus\dots\oplus so(l_p),
\end{align*}
and let $\mathfrak v$ be the orthogonal complement with respect to the invariant scalar product
$\langle\cdot,\cdot\rangle$.
Then the inertia operator \eqref{MANAKOV} preserves the decomposition $so(n)=so(l_1)\oplus so(l_2)\oplus\dots\oplus so(l_p)\oplus\mathfrak v$:
\[
I=2\alpha_1\pr_{so(l_1)}+\dots+2\alpha_p\pr_{so(l_p)}+I\circ\pr_{\mathfrak v}.
\]

Using a slight modification of the Manakov Lax representation, we have the following theorem.

\begin{thm}\label{manakov-giroskop}
(i) The equations of motion of a rigid body with a gyroscope with the angular momentum $L\in\mathfrak h$ imply the
matrix polynomial Lax representation
\begin{equation}
\dot {\mathbb L}(\lambda)=[{\mathbb L}(\lambda), {\mathbb A}(\lambda)],\qquad {\mathbb L}(\lambda)=M+L+{\lambda} J^2,\qquad {\mathbb A}(\lambda)=\Omega+\lambda J.
\label{matrix}
\end{equation}

(ii) For a generic $L\in\mathfrak h$, the symmetric Manakov top with a gyroscope is Liouville integrable.
\end{thm}

\noindent\emph{Proof.}
As in \cite{Ma},  item (i) follows from the relation
\[
[J^2,\Omega]+[M,J]=[J^2,\Omega]+[J\Omega+\Omega J,J]=0,
\]
together with the property $[L,J]=0$.
The rest of the proof is the same as the proof of item (ii) in Theorem \ref{glavna}, based on  Theorem 1 in \cite{DGJ2009}.
Namely, set $M+L=K$ and consider $so(n)^*(K)$ with the Lie-Poisson bracket
\[
\{K_{ij}, K_{jk}\}_0=-K_{ik}.
\]
Let $\mathcal S$ be the set of linear functions on $\mathfrak h=so(n)_J$.
The Euler equations of a symmetric free rigid body
\[
\dot K=[K,\Omega], \qquad \Omega=I^{-1}(K),
\]
have a set of Noether first integrals $\mathcal S$ together  with the set of the Manakov first integrals:
\[
\mathcal L=\Big\{\tr\big((K+{\lambda} J^2)^{2k}\big)\,\vert\, \lambda\in\R, \, k=1,\dots,\Big[\frac{n}2\Big]\Big\}.
\]
Then, $\{\mathcal S,\mathcal L\}_0=0$, $\{\mathcal L,\mathcal L\}_0=0$ and
$\mathcal S + \mathcal L$ is a complete set of functions on $so(n)^*$:
\[
\ddim(\mathcal S+\mathcal L)+\dind(\mathcal S+\mathcal L)=\dim(so(n))+\corank(\{\cdot,\cdot\}_0)=\dim(so(n))+\rank(so(n)).
\]
This provides non-commutative integrability of a free symmetric rigid body with the
 Hamiltonian function $H_0=\frac12\langle I^{-1}K,K\rangle$ and
 the mass tensor defined by \eqref{symJ} (see \cite{DGJ2009}).  Now, the integrability of  the system with the Hamiltonian
$H_1=H_0-\langle K,I^{-1} L\rangle$ follows applying the same arguments as  in the proof of Theorem \ref{glavna} -- by constructing a complete commutative subset of  first integrals from the set of polynomials $\mathcal S$ using the Mishchenko-Fomenko argument translation method.
\hfill $\Box$

\subsection*{Acknowledgements}
 This research has been supported by the grant IntegraRS of the Science Fund
of Serbia, the Ministry for Education, Science, and Technological Development of Serbia, and the Simons
Foundation grant no. 854861.

\end{document}